\documentclass[11pt]{benc}
\usepackage{graphicx}
\usepackage{times}
\usepackage{textcomp,gensymb}
\usepackage{units}
\usepackage{fancyheadings}
\usepackage[amsmath]{benc}

\newcommand{\nitride}{Si\tsub{3}N\tsub{4}}

\begin{document}

\title{Microfabrication of Laser-Driven Accelerator Structures%
\thanks{Work supported by Department of Energy
contract DE-AC03-76SF00515 (SLAC) and by DOE grant
no. DE-FG03-97ER41043-II.}}
\author{
B. Cowan \\
\medskip
{\small
Stanford Linear Accelerator Center, Stanford University,
Stanford, California 94309}
}
\date{}

\maketitle

\thispagestyle{fancy}
\setlength{\headrulewidth}{0pt}
\rhead{
SLAC--PUB--9704 \\
August 2002
}

\begin{abstract}
We discuss the potential for using microfabrication techniques for
laser-driven accelerator construction.  We introduce microfabrication
processes in general, and then describe our investigation of a
particular trial process.  We conclude by considering the issues
microfabrication raises for possible future structures.

\begin{center}
\emph{Submitted to Tenth Advanced Accelerator Concepts Workshop (AAC 2002),
June 23--28, 2002, Mandalay Beach, California (AIP Conference Proceedings)}
\end{center}
\end{abstract}

\section{Introduction}

Experiments to demonstrate laser-driven charged particle acceleration
in vacuum have thus far relied largely on macroscopic structures, that is,
structures large enough to be assembed by hand and many optical
wavelengths in size.  For instance, in the LEAP experiment conducted
at Stanford University \cite{leap1}, the accelerator cell is
approximately \unit[1]{cm} long in each dimension compared to a laser
wavelength of \unit[800]{nm}, and is assembled by manually gluing
together high-reflection coated quartz prisms.

Such accelerator structures are not designed to be candidates for use
in a real accelerator, and indeed their capabilities are clearly not
close to what would be needed in a functioning machine.  For
instance, the LEAP cell has a maximum accelerating gradient of
\unit[10]{MeV/m}, and with acceleration of \unit[1]{pC} bunches over
a distance of \unit[1.5]{mm} using \unit[36]{\micro J} laser pulses at
the LEAP cell, the structure has an electron-to-photon efficiency of
$4.2\e{-4}$.  With just one cell, the overall wall-plug efficiency of
the setup is much less, close to $10^{-10}$.

These performance parameters can be traced fundamentally to the large size
of the structure; for effective accelerator cells a much smaller
structure is required.  Consider a laser field propagating in free
space, which largely describes the field in the LEAP cell.  The peak
longitudinal field $E_z$ on axis obeys the general scaling law
$E_z/E_x\sim\lambda/w_0$, where $E_x$ is the peak transverse field,
$\lambda$ is the wavelength, and $w_0$ is the transverse mode size.
This relation is exact for the case of a Gaussian TEM\tsub{10}
mode, but even approximately this scaling law serves to illustrate the
dependence of $E_z$ on the mode size.  At the same time, the maximum
field amplitude $\abs{\vect{E}}$ is fixed by the damage threshold of
the optics, so higher gradients cannot be obtained by arbitrarily
increasing the laser intensity.  Therefore, for optimum gradient the
laser mode size must be comparable to the wavelength, as is the case
with RF structures.

Attaining good shunt impedance in a laser-driven structure also
constrains possible designs.  For reasonable efficiency a structure
must accelerate particles continuously over a distance large compared
to a wavelength.  Therefore simply focusing a low-intensity laser
pulse down to a very narrow waist to avoid optical damage will not be
effective, since the small spot will diffract quickly and will not
accelerate for any appreciable distance.  Also, a structure may need
to store laser energy for the next electron bunch.  Therefore, a
structure with both high gradient and good efficiency must be small, with
feature sizes on the order of an optical wavelength.  Fortunately,
technology now exists or is rapidly being developed to fabricate such small
structures, even for wavelengths in the visible or near-infrared.

\section{Microfabrication Overview}
We refer to microfabrication, in general, as a type of process using a
set of techniques and equipment commonly used to manufacture
integrated circuits (IC's) and microelectromechanical devices (MEMS).
Such a process typically starts with a bare wafer, usually of
silicon.  Processing of the wafer involves several types of
procedures, which we describe briefly below; detailed discussions can
be found in references on IC or MEMS techniques, for instance
\cite{plummer} or \cite{madou}.

\emph{Photolithography}, perhaps the most important process step,
defines the structure by transferring a pattern from a mask onto an
organic photoresist on the wafer surface.  This is usually followed by
\emph{etching}, in which material is selectively removed from the
wafer.  \emph{Thin film deposition} procedures allow many different
materials to be deposited in layers on the wafer, and a thin film of
SiO\tsub{2} can be grown directly on a silicon surface by the process of
\emph{thermal oxidation}.  \emph{Ion implantation} allows dopants to
be added to the structure, which diffuse through the material during
subsequent high-temperature process steps.  In IC manufacturing this
is primarily used to control electrical properties of materials, but
it can be used to control chemical and possibly optical properties as
well.  \emph{Chemical-mechanical polishing} has recently become a
common procedure for planarizing the topography of a wafer at certain
points in a process.  Finally, the importance of \emph{cleaning}
cannot be ignored.  Usually involving immersion in chemical baths,
each followed by a deionized water rinse, as well as adherence to
cleanroom procedures, removing particulates and chemical contaminants
to suppress defects is critical to IC manufacturing and will probably
be equally important in accererator microstructure fabrication.

Microfabrication has great potential for accurately making small
accelerator structures.  Not only are highly precise procedures
available now, but continued rapid improvement is driven by the
extraordinary market forces in the IC industry.  For instance, control
of certain feature sizes is expected to reach \unit[0.5]{nm} RMS by
2010 \cite{ITRS01}.
That silicon processes are so well established
can be exploited for optically-driven accelerators, since silicon
transmits in the telecommunications band at \unit[1.5]{\micro m}
wavelength, where many optical components are available and continue
to be improved.  For research purposes, there is an advanced
fabrication facility at Stanford \cite{SNF} where the procedures
mentioned above are available.  Finally, subsequent mass production
using microfabrication techniques is inexpensive; this is critical since,
for instance, an accelerator the length of the SLAC linac would use
10,000 300-mm wafers, the current state-of-the-art wafer size.

\section{Trial Process}
\begin{figure}
\begin{center}
\resizebox{.8\textwidth}{!}{\includegraphics{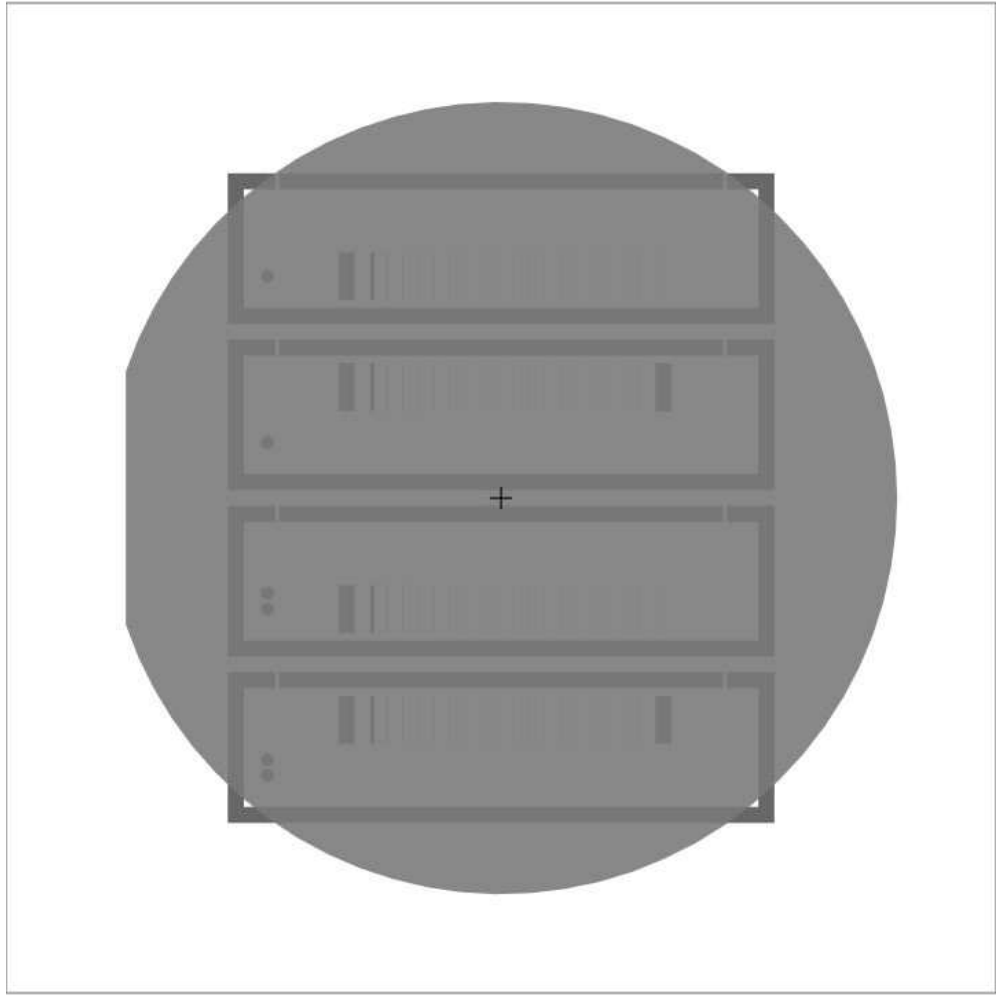}}
\caption{The mask pattern for the trial process.  The dark area
corresponds to the region to be etched.  The wafer flat is in a
$\{111\}$ crystal plane.  T. Plettner contributed to this figure.}
\label{fig:mask}
\end{center}
\end{figure}

We have explored a trial microfabrication process for a replacement
LEAP cell.  While the purpose of this project was primarily to explore
microfabrication procedures, it was motivated by the current LEAP cell
design.  As described in \cite{leap1}, the LEAP accelerator structure
consists of two pairs of dielectric surfaces oriented at 45\degree\ to
each other, with a slit in the middle for electron beam passage.  Each
surface has a high-reflector dielectric coating.  The microfabricated
cell is designed to form the e-beam slits by etching them through a
silicon wafer.  The etch pattern on the wafer is shown in
Figure~\ref{fig:mask}.
Up to four reflecting rectangles would be etched from a 4-inch wafer,
and each surface would have several slits.  The rectangles would then
be aligned in pairs at a 45\degree\ angle.  Unlike the current LEAP
cell, the slits would not be individually adjustable, but there would
be slits of several widths in each rectangle.  In addition, having
several slits would allow a new slit to be moved into place easily
should one be damaged during the experiment.

The process of etching a pattern in a silicon wafer can be summarized
simply as follows: We deposit layers of material on the wafer, and
then selectively remove the areas to be etched from each layer, with
each deposited layer protecting the regions not to be etched of the
layer beneath.  However, each step of the process involves a different
procedure using different equipment.  We now describe in some detail
each step of the process.

The final step of the process will be to etch the silicon in a
potassium hydroxide (KOH) solution.  While the pattern will initially
be imprinted in photoresist, the resist cannot be used directly to
mask the KOH since the KOH will eat it away as it etches the silicon.
Therefore an intermediate layer is required to mask the silicon.
Silicon nitride (\nitride) provides a good masking material, as it
etches negligibly slowly in KOH \cite{bean}.  Therefore, the first
step of the process is to deposit a thin film of \nitride\ on the
silicon substrate.

We deposit a \unit[200]{nm} film of \nitride using low-pressure
chemical vapor deposition (LPCVD).  In this procedure dichlorosilane
and ammonia gases are run through a furnace containing the wafers at
about \unit[1]{torr} and 700\celsius.  These gases react on
the wafer surface to produce \nitride.  One hour of deposition is
sufficient to give a \unit[200]{nm} film.

One the nitride is deposited we spin on a layer of photoresist.  This
is accomplished by dropping a small puddle of the liquid resist
compound on the wafer and then spinning the wafer at 3000--5000 RPM
for about \unit[60]{s}.  After the first few seconds of spinning the
resist coats the wafer uniformly; the rest of the spinning serves to
dry the resist.  The resist is then baked to harden it to prevent flow
during the development step, discussed below.

\begin{figure}
\begin{center}
\resizebox{0.8\textwidth}{!}{\includegraphics{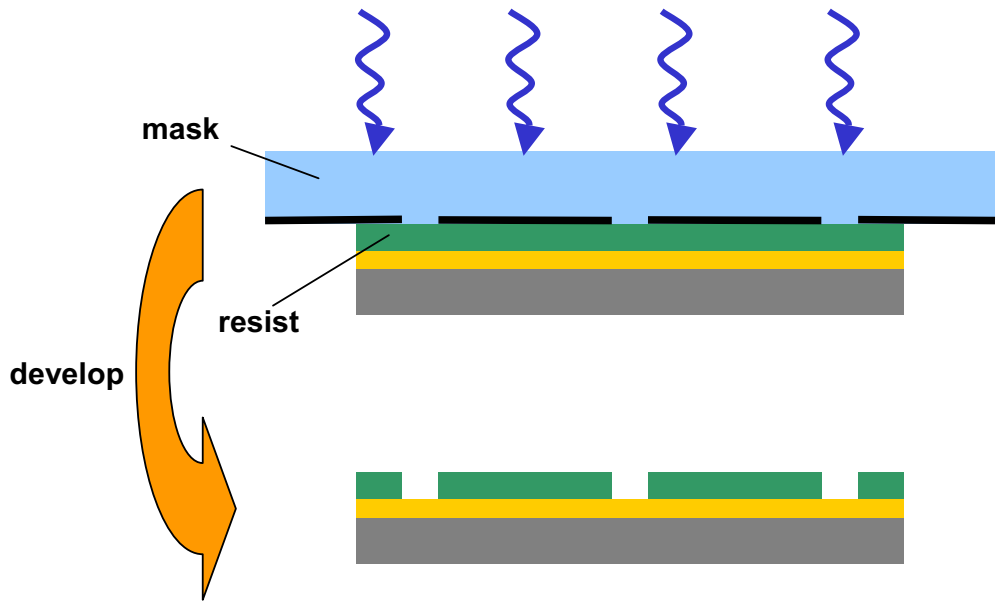}}
\caption{Patterning the photoresist.}
\label{fig:pattern}
\end{center}
\end{figure}
The pattern is transferred to the photoresist by placing a
chome-on-glass mask over the wafer, with chrome absent from regions
corresponding to etched regions on the wafer.
The wafer is then brought into contact with the mask and the mask is
exposed from above to 365-nm UV light from a mercury lamp.  This
exposes only the regions of photoresist to be removed, causing a
chemical reaction in the resist.  The resist compound was deliberately
chosen to be thin to avoid diffraction effects; this is a concern
because the slits in the mask are as thin as \unit[1]{\micro m}.  After
a bake, the wafer is immersed in a developer solution, removing the
exposed regions of photoresist.  This procedure is shown schematically
in Figure~\ref{fig:pattern}.

Once the photoresist is patterned, the wafer is plasma
etched using CF\tsub{4} and O\tsub{2} gases.  This removes
the nitride not protected by photoresist.  Once the plasma etch is
complete, the resist can be removed using chemical solvents plus a
quick plasma etch in O\tsub{2}, oten called a plasma ``ash,'' to
remove any remaining organic residues.

\begin{figure}
\begin{center}
\resizebox{.8\textwidth}{!}{\includegraphics{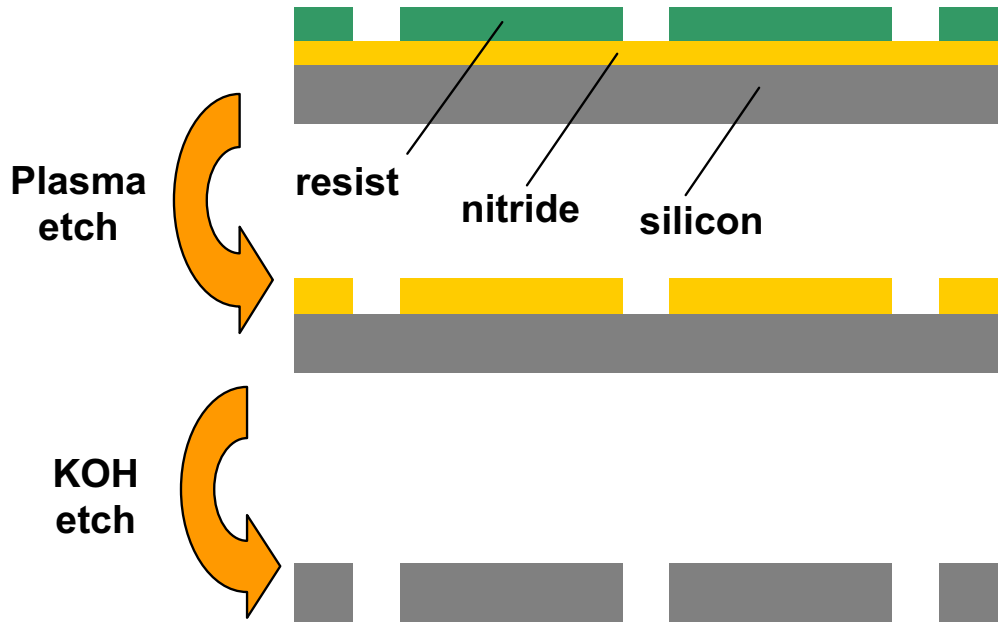}}
\caption{Etching the slit pattern in the wafer.}
\label{fig:etch}
\end{center}
\end{figure}

Finally, the wafer is etched in KOH solution.  The etch
is highly anisotropic, proceeding much faster in the $\ip{110}$
direction than in the $\ip{111}$ direction.  In fact, etch ratios of
greater than 600 to 1 have been obtained \cite{bean}.  This is why the wafer
flat in Figure~\ref{fig:mask} is oriented in the $\{111\}$ plane,
and it also requires that the wafer surface be a $\{110\}$ plane.  We
have found that ultrasonically agitating the KOH solution improves
the etch rate considerably.  Once the KOH etches entirely through
the wafer, the nitride can be removed using the same plasma etch used
above or in a hot phosphoric acid solution, and then the optical
coating can be applied.  The etch process is shown schematically in
Figure~\ref{fig:etch}.

So far, we have obtained a \unit[1]{\micro m} slit in the nitride
coating, and etched all the way through a
\unit[500]{\micro m}-thick wafer.  However, the narrowest slit we
have been able to obtain is still \unit[40--50]{\micro m} wide,
and better alignment between the mask and the $\{111\}$ crystal
plane will likely be able to improve upon this.  The slit walls appear
vertical under an optical microscope but proper observation of the
structure requires an SEM scan.  Depositing the dielectric coating
without clogging the slits may prove difficult, especially because
high-reflector coatings require many layers.  As described in
\cite{plummer}, thin film deposition geometry is highly nontrivial and
an appropriate simulation would be required.

\section{Microfabrication Possibilities}

The use of microfabrication presents possibilities for laser-driven
accelerator structures well beyond those available through bulk
machining.  One possibility is to use a photonic crystal structure,
for instance as proposed by Lin \cite{lin}.  However, the available
procedures in a microfabrication process do impose constraints on the
type of structure used.

The structure material must be lossless at a convenient laser
wavelength, radiation hard, and the substrate material must have high
thermal conductivity.  Also, the structure material must be easily
etched, and if the structure is made of more than one material there
must be processes available to selectively etch them independently.
Therefore a
glass photonic crystal fiber as proposed in \cite{lin} is not viable
because SiO\tsub{2} is not radiation hard, and fibers do not release heat
easily.

However, other materials may be used for a similar
structure.  Silicon, as mentioned above, transmits at the
\unit[1.5]{\micro m} wavelength, and is possibly is even better
suited for use at a \unit[2.5]{\micro m} wavelength.  There is an
immense microfabrication technology base for silicon, and it may be
used as a substrate, even if it is not suited for a structure, because
of its wide availability and good thermal conductivity.  Quartz and
sapphire are also possible materials, as is diamond.  Photonic crystal
structures might use these materials, and in fact a mid-infrared
photonic bandgap structure has been microfabricated in silicon
\cite{roddle}.

It is worth noting that there are several differences in fabrication
considerations between IC and optical structure fabrication, and that
the task of optical structure fabrication may in fact be easier than
IC manufacturing for these reasons.  For instance, laser-driven
accelerator structures are unlikely to depend on particular doping
profiles, whereas maintaining precise dopant concentrations is critical in the
IC industry.  Therefore IC manufacturers have a limited thermal budget
for their processes, while we are free to use high-temperature
procedues at will during an accelerator structure process.  Also, IC's
involve quite a few materials of different chemical properties and
many mask geometries.  By contrast, accelerator structures will likely
have much greater symmetry or regularity.  In fact, the structure described in
\cite{roddle} uses only one material and just one mask in a repetitive
process which is quite simple compared to a typical IC manufacturing
process.

All of these considerations give microfabrication techniques great
potential for producing laser-driven accelerator structures.  The
variety of techniques and equipment is vast, and we look forward to
exploring further their use as we investigate possible accelerator
structures.

\section{Acknowledgements}

Thanks to J. Mansell for valuable input on the trial process and
N. Latta, M. Mansourpour, and U. Thumser of SNF for helpful guidance
on using the equipment.  Work supported in part by Department of Energy
contract DE-AC03-76SF00515 (SLAC) and by DOE grant
no. DE-FG03-97ER41043-II.


\begin{thebibliography}{99}
\bibitem{leap1} T. Plettner et. al., ``Progress of the Laser
Electron Accelerator Project at Stanford University,'' in
\emph{Proceedings of the 2001 Particle Accelerator Conference}, edited
by P. Lucas and S. Webber, pp. 108--110
\bibitem{plummer} Plummer, J. D., Deal, M. D., and Griffin, P. B., 
\emph{Silicon VLSI Technology: Fundamentals, Practice, and
Modeling}, Prentice Hall, 2000
\bibitem{madou} Madou, M. J., \emph{Fundamentals of Microfabrication:
The Science of Miniaturization}, 2nd ed., CRC Press, 2002
\bibitem{ITRS01} ``International Technology Roadmap for
Semiconductors'', SIA 2001.  See \texttt{http://www.semi.org/}.
\bibitem{SNF} The Stanford Nanofabrication Facility; see
\texttt{http://snf.stanford.edu}.
\bibitem{bean} Bean, K. E., \emph{IEEE Trans. Electron. Devices}
\textbf{25} 1185--93 (1978)
\bibitem{lin} Lin, X. E., \emph{Phys. Rev. Special Topics,
Accelerators and Beams} \textbf{4} 051301 (2001)
\bibitem{roddle} S. Y. Lin et. al., \emph{Nature} \textbf{394}
251--253 (1998)
\end{thebibliography}
\end{document}